\documentclass[aps,prl,superscriptaddress,twocolumn,floatfix]{revtex4-1} 

\usepackage{amsmath}  
\usepackage{amsfonts} 
\usepackage{graphicx} 
\usepackage{newfloat} 
\usepackage{gensymb} 
\usepackage{float} 
\usepackage{accents}

\usepackage[pdftex,bookmarks=true,bookmarksopen,bookmarksnumbered,colorlinks,linkcolor=blue,citecolor=blue]{hyperref}

\renewcommand{\figurename}{Figure}
\makeatletter
\renewcommand*{\fnum@figure}{{\normalfont\bfseries \figurename~\thefigure}}
\renewcommand*{\@caption@fignum@sep}{ $|~$}

\DeclareFloatingEnvironment[
	fileext=los,
	name=Figure,
	placement=tbhp,
	within=none,
]{suppfig}

\renewcommand{\suppfigname}{Supplementary Figure}
\makeatletter
\renewcommand\thesuppfig{S\arabic{suppfig}}
\renewcommand*{\fnum@suppfig}{{\normalfont\bfseries \suppfigname~\thesuppfig}}

\newcommand{\beginsupplement}{%
	\setcounter{table}{0}
	\renewcommand{\thetable}{S\arabic{table}}%
	\setcounter{figure}{0}
    \pagebreak
    \widetext
}

\begin{document}


\title{Single-electron operation of a silicon-CMOS 2x2 quantum dot array with integrated charge sensing}

    \author{Will Gilbert}
    \email[w.gilbert@unsw.edu.au]{}
	\affiliation{School of Electrical Engineering and Telecommunications, The University of New South Wales, Sydney, NSW 2052, Australia.}
	\author{Andre Saraiva}
	\email[a.saraiva@unsw.edu.au]{}
	\affiliation{School of Electrical Engineering and Telecommunications, The University of New South Wales, Sydney, NSW 2052, Australia.}
	\author{Wee Han Lim}
	\affiliation{School of Electrical Engineering and Telecommunications, The University of New South Wales, Sydney, NSW 2052, Australia.}
	\author{Chih Hwan Yang}
	\affiliation{School of Electrical Engineering and Telecommunications, The University of New South Wales, Sydney, NSW 2052, Australia.}
	\author{Arne Laucht}
	\affiliation{School of Electrical Engineering and Telecommunications, The University of New South Wales, Sydney, NSW 2052, Australia.}
	\author{Benoit Bertrand}
	\affiliation{Universit\'e Grenoble Alpes, CEA, LETI, 38000 Grenoble, France.}
	\author{Nils Rambal}
	\affiliation{Universit\'e Grenoble Alpes, CEA, LETI, 38000 Grenoble, France.}
	\author{Louis Hutin}
	\affiliation{Universit\'e Grenoble Alpes, CEA, LETI, 38000 Grenoble, France.}
	\author{Christopher C. Escott}
	\affiliation{School of Electrical Engineering and Telecommunications, The University of New South Wales, Sydney, NSW 2052, Australia.}
	\author{Maud Vinet}
	\affiliation{Universit\'e Grenoble Alpes, CEA, LETI, 38000 Grenoble, France.}
	\author{Andrew S. Dzurak}
	\email[a.dzurak@unsw.edu.au]{}
	\affiliation{School of Electrical Engineering and Telecommunications, The University of New South Wales, Sydney, NSW 2052, Australia.}
	
\begin{abstract}
The advanced nanoscale integration available in silicon complementary metal-oxide-semiconductor (CMOS) technology provides a key motivation for its use in spin-based quantum computing applications. Initial demonstrations of quantum dot formation and spin blockade in CMOS foundry-compatible devices are encouraging, but results are yet to match the control of individual electrons demonstrated in university-fabricated multi-gate designs. We show here that the charge state of quantum dots formed in a CMOS nanowire device can be sensed by using floating gates to electrostatically couple it to a remote single electron transistor (SET) formed in an adjacent nanowire. By biasing the nanowire and gates of the remote SET with respect to the nanowire hosting the quantum dots, we controllably form ancillary quantum dots under the floating gates, thus enabling the demonstration of independent control over charge transitions in a quadruple ($2\times2$) quantum dot array. This device overcomes the limitations associated with measurements based on tunnelling transport through the dots and permits the sensing of all charge transitions, down to the last electron in each dot. We use effective mass theory to investigate the necessary optimization of the device parameters in order to achieve the tunnel rates required for spin-based quantum computation.
\end{abstract}

\date{\today}
\maketitle

\section*{INTRODUCTION}

Recent progress in the development of silicon-based spin qubit devices has pushed the state-of-the-art close to the minimum requirements for fault-tolerant quantum computing. High fidelity one- and two-qubit operations have been demonstrated using prototype devices, fabricated in university-based research facilities \cite{yoneda2018quantum, yang2019silicon, xue2019benchmarking, zajac2018resonantly, huang2019fidelity}. In particular for silicon-MOS quantum dot qubits, their similarity to the ubiquitous MOSFET significantly increases their potential as a platform for full-scale quantum computation. Given the reproducibility and yield provided by current industrial nanofabrication standards of the CMOS industry \cite{vinet2018towards, pillarisetty2019high}, it is possible to envisage integrated quantum dot devices reaching the millions of qubits needed for quantum error correction protocols \cite{vandersypen2017interfacing}. Despite this, research into the use of foundry-produced CMOS devices for spin-based qubits has only recently been explored \cite{maurand2016cmos, corna2018electrically}. This is, at least in part, because fabricating qubit prototype devices in a CMOS foundry often requires layouts not fully  compliant with the conventional design rules for industrial processes. 

Here we employ a foundry-made quantum dot, formed in a silicon nanowire, as a single electron transistor (SET), and use this for charge sensing \cite{podd2010charge} of a remote, multi-quantum dot device formed in an adjacent nanowire. The remote sensing is facilitated by floating gates (couplers) \cite{buehler2006controlled}, which couple the sensor nanowire and the quantum dot nanowire, providing enhanced sensitivity of charge transitions down to the last electron. Furthermore, by biasing all of the gates, together with the source and drain, on the sensor nanowire with respect to the dot nanowire, the couplers themselves accumulate additional quantum dots such that we are able to form and sense a quadruple ($2\times2$) quantum dot array, operated in the single electron regime. Progress in sensing foundry-fabricated quantum dot arrays has been made recently in devices where the sensor is integrated immediately next to the quantum dots \cite{ansaloni2020single, chanrion2020charge}. In our work, the remote charge sensor with floating couplers allows for detection of both dot-reservoir and interdot charge transitions with high sensitivity, whilst reducing the layout impact of the sensor. In a two-dimensional qubit array, floating couplers enable remote and flexible positioning of the sensor.

\begin{figure*} [ht]
	\centering
	\includegraphics[width=1\linewidth]{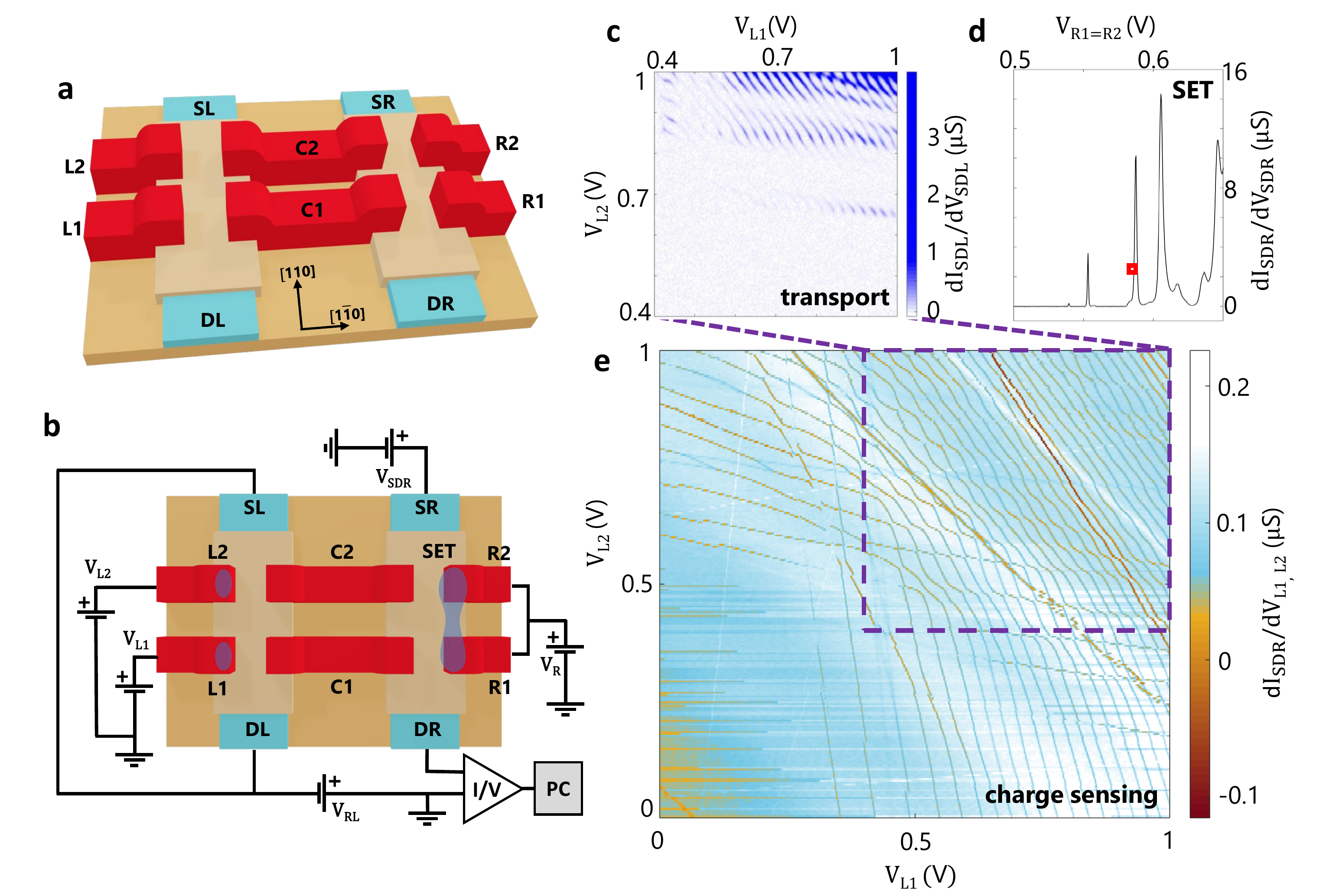}
	\caption{\textbf{Device architecture and charge sensing configuration.} 
		\textbf{a}, Three-dimensional schematic layout of the device, showing two parallel FD-SOI silicon nanowires connected to independent sources (SL and SR) and drains (DL and DR). The two nanowires have a high-quality gate oxide and their electrostatic potential is controlled by the titanium nitride and polysilicon gates L1 and L2 on the left and R1 and R2 on the right. Furthermore, two floating gates C1 and C2 overlap the opposite corners of both nanowires, enhancing the electrostatic coupling between the nanowires. \textbf{b}, Measurement schematic also showing simultaneous formation of a double dot under L1 and L2 and a large dot in the right nanowire spanning both R1 and R2. \textbf{c}, Conductance dI$_{\mathrm{SDL}}$/dV$_{\mathrm{SDL}}$ as a function of the bias on gates L1 and L2 reveals current peaks below the threshold voltage indicating the formation of quantum dots. The typical honeycomb shape of this transport map indicates that two dots form under L1 and L2. As the gate biases are lowered, the tunnel coupling from the dots to source and drain becomes small and no current is observed. \textbf{d}, Conductance of the right nanowire dI$_{\mathrm{SDR}}$/dV$_{\mathrm{SDR}}$ reveals subthreshold Coulomb oscillations indicating dot formation. The red square indicates a point of optimal sensitivity of the current I$_{\mathrm{SDR}}$ to changes in the dot chemical potential, which we use for charge sensing of the dots in the left nanowire. The charge sensor is kept at its highest sensitivity configuration using a feedback system. \textbf{e}, Charge sensing signal of L1 and L2 dots. Changes in the charge distribution in the left nanowire impact the chemical potential of the large dot in the right nanowire through the floating electrostatic couplers. Tracking the charge transitions through remote charge sensing in the right nanowire leads to enhanced sensitivity to charge transitions when the dot is depleted to the few-electron regime compared to transport spectroscopy (for reference, the voltage space scanned in panel \textbf{c} is marked by the dashed square line). This allows us to monitor the quantum dot system down to the last electron.
	}\label{device}
\end{figure*}

\section*{Device design and characterisation}

The device architecture used in this work is shown in Figure~\ref{device}a and consists of two parallel 70~nm wide silicon nanowires fabricated using fully-depleted silicon-on-insulator (FD-SOI) technology at CEA-Leti \cite{hutin2016si}. The nanowires are 7~nm thick and have an effective oxide thickness of 6~nm. The gate electrodes L1, L2, R1, R2, made out of a stack of titanium nitride and polysilicon, wrap over the edges of the nanowires and are connected to DC voltage sources for electrostatic control (details in Figure~\ref{device}b). Additional uncontacted gates C1 and C2 serve as floating couplers between the nanowires. Silicon nitride spacers are used to self-align the n-type doped source (SR/SL) and drain (DR/DL) leads to the gates.

The left-side nanowire gates L1 and L2 were individually biased to accumulate one quantum dot under each gate. This two-dot system was first characterised via current (transport) measurements \cite{van2002electron} as a function of bias voltages V$_{\mathrm{L1}}$ and V$_{\mathrm{L2}}$, shown in Figure~\ref{device}c. The honeycomb pattern associated with coupled quantum dots can be resolved only for a high number of electrons in each dot (large V$_{\mathrm{L1}}$ and V$_{\mathrm{L2}}$). For low electron occupancies, the smaller electronic wavefunctions result in reduced tunnel rates and immeasurably low currents for this measurement set-up.

The right-side nanowire was then operated as an SET charge sensor by biasing the (electrically shorted) R1 and R2 gates to accumulate one quantum dot, which serves as the SET island, coupled to heavily-doped source and drain regions SR and DR. Large amplitude Coulomb blockade oscillations in the sensor nanowire current are observed (Figure~\ref{device}d). The irregular Coulomb blockade oscillations observed in this device can be corrected with independent control of R1 and R2 bias (see Supplementary Information). Biasing the SET to the side of a Coulomb blockade peak, as marked by the square symbol in Figure~\ref{device}b, makes the SET conductance sensitive to nearby changes in the potential \cite{podd2010charge}. In doing so, the right-side nanowire SET charge sensor can detect charge movement in the left-side dot nanowire, mediated by the floating electrostatic couplers and providing charge transition sensitivity down to the last electron for quantum dots under L1 and L2, as can be seen in Figure~\ref{device}e. As the charge occupancy of the dots is increased with more positive bias V$_{\mathrm{L1}}$ and V$_{\mathrm{L2}}$, the inter-dot coupling increases as indicated by the emergence of a traditional honeycomb pattern \cite{van2002electron}. At higher gate biases the transitions eventually become diagonal lines, corresponding to a completely merged large dot. Decreased visibility of certain charge transitions in the few-electron regime is seen where the tunnel rate between the quantum dots and the reservoir becomes slower than the lock-in probe frequency \cite{yang2011dynamically} (213~Hz in this case). Additional transition lines seen in Figure~\ref{device}e may correspond to disorder or randomly positioned dopants. Dashed lines mark the full window in which transport measurements were taken for Figure~\ref{device}c and demonstrate the superiority of this charge-sensing arrangement for the observation of charge transitions in the double dot system  \cite{nordberg2009charge}.

The electrostatic coupling between the sensor nanowire and the quantum dots is enhanced by the presence of the floating gates C1 and C2 \cite{buehler2006controlled}. A similar double nanowire device in which no floating coupler is present, so that the dot gates wrap over their respective nanowires, was also measured, revealing a reduced sensitivity (see Supplementary Material).

\section*{OPERATION OF A 2x2 QUANTUM DOT ARRAY} 

The dimensionality of qubit arrangements plays an important role in the propagation of errors, as well as in the fidelity threshold for quantum error correcting codes. A non-trivial topology for quantum dot networks is therefore a key development towards full scale silicon quantum computers. Even though the device design investigated here does not allow for an extended two-dimensional arrangement of qubits, we take a first step towards this goal by developing a technique to accumulate and characterise a $2\times2$ array of dots within a single nanowire.

\begin{figure*} [ht]
	\centering
	\includegraphics[width=\linewidth]{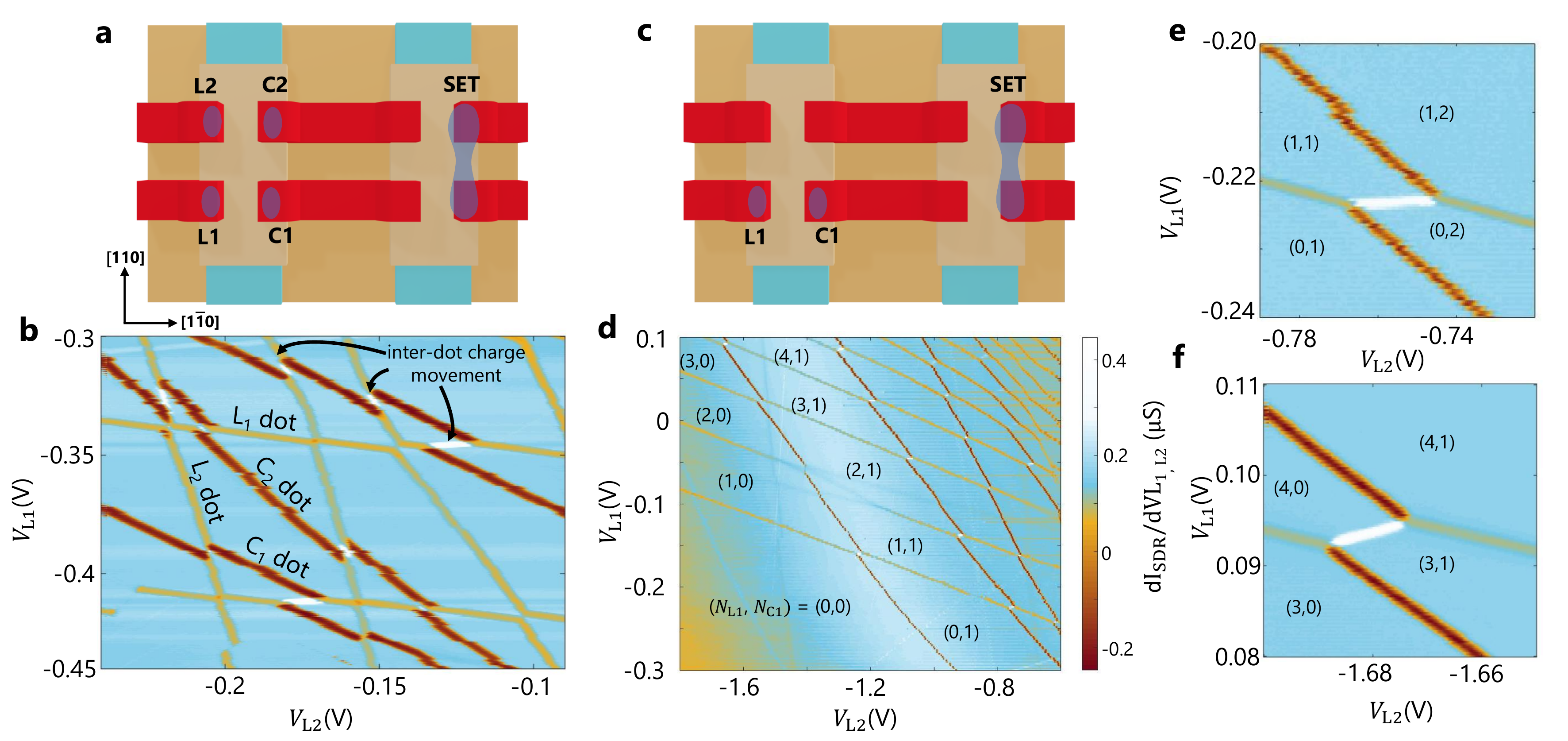}
	\caption{\textbf{Formation of dots under the floating couplers and charge configurations of interest for quantum computation.}
	\textbf{a}, Schematics of the formation of a $2\times2$ array of quantum dots achieved by biasing all gates and source and drain in the right nanowire with respect to those in the left nanowire. The induced potential of the coupler gates C1 and C2 leads to the accumulation of two additional dots in the left nanowire. \textbf{b}, Charge sensing spectroscopy of the quadruple dot. All charge transitions are distinguishable due to their different electrostatic couplings to gates L1 and L2 and to the charge sensor, leading to different slopes and contrasts in the colour plot (colour scales are the same in panels \textbf{b} and \textbf{d}). The transitions at more vertical and horizontal angles with less contrast refer to changes in the number of electrons at the dots under L1 (horizontal) and L2 (vertical). The transitions at intermediate angles with higher contrast correspond to transitions in the dots forming under C1 and C2. White regions are associated with the movement of charges between an L dot and a C dot. \textbf{c}, Schematics of a double quantum dot formed by depleting the dots L2 and C2 completely and having few electrons under L1 and C1. \textbf{d}, The charge stability diagram of this configuration indicates that a well-defined double quantum dot is formed with distinguishable charge configurations for various dot occupancies of interest for quantum computation. \textbf{e,f}, Details of the charge transitions for (1,1)-(0,2) and (4,0)-(3,1), respectively.}\label{quadruple}
\end{figure*}

The strong electrostatic coupling between the floating gates (C1 and C2) and each nanowire allows the creation of additional quantum dots underneath the floating gates by using a differential bias between the nanowires. The left-side (dot) nanowire was biased negatively relative to the right-side (sensor) nanowire, such that the floating couplers C1 and C2 act as gates to induce two additional quantum dots, shown schematically in Figure~\ref{quadruple}a. Measurements with this configuration are shown in Figure~\ref{quadruple}b. Charge transitions in C1 and C2 dots are distinguishable from L1 and L2 quantum dot transitions on the V$_{\mathrm{L1}}$ vs V$_{\mathrm{L2}}$ charge stability diagram by nature of their relative coupling strengths to the gates L1 and L2. The L1 and L2 quantum dots have strong coupling to the biased gates and therefore exhibit transitions nearly perpendicular to their respective axes. The quantum dots under C1 and C2 have larger cross capacitance to the opposite dot gate (i.e. L2 and L1, respectively), leading to transitions that are inclined with respect to the axes. Secondly, the C1 and C2 quantum dots couple more strongly to the sensor compared to L1 and L2 dots, owing to their proximity to the floating couplers, and hence their transitions can be distinguished by the larger influence on the SET current (see Figure~\ref{quadruple}b). 

This arrangement of quantum dots provides increased connectivity and dimensionality while also providing a configuration favourable for single-shot readout of inter-dot transitions, which is a requirement for qubit readout based on Pauli spin blockade. We note that similar $2\times2$ quantum dot arrays were formed in devices fabricated with the same foundry technology, but which contain a single nanowire and all the split gates are directly biased (no floating coupler gate) \cite{ansaloni2020single, chanrion2020charge}. In contrast to that strategy, our method enables an SET to be positioned remotely to the quantum dot array, while maintaining a high sensitivity to interdot charge movements, which can be seen in Figure~\ref{quadruple}b as white lines. Transitions between L1 and L2 have less signal since the charge movement occurs parallel to the SET nanowire. We focus now on a double dot configuration with electrons under L1 and C1, as schematically depicted in Figure~\ref{quadruple}c. The corresponding charge transition diagram in Figure~\ref{quadruple}d shows operation in the few electron regime.

For quantum computation, the quantum level the electron occupies in either dot should be well separated energetically from excited states. If this is not the case, spin-triplet states can form and thereby prevent Pauli spin blockade. Due to the low symmetry of these corner dots, a detailed analysis of the electron filling and assigning quantum numbers to the different dot occupations is a challenging task which is beyond the scope of the present work. Instead, we focus on the charge configurations in which the electrons occupy either the ground valley-orbital state (Figure~\ref{quadruple}e); or the configuration in which two electrons could form a closed shell in the ground valley-orbital state and a third electron occupies an excited valley state (Figure~\ref{quadruple}f). In what follows, all charge occupation values (N$_\mathrm{L1}$, N$_\mathrm{C1}$) refer to the configuration with N$_\mathrm{L1}$ electrons under the L1 dot and N$_\mathrm{C1}$ electrons under the C1 dot.

\section*{TUNNEL RATE MEASUREMENT}

All possible spin-based quantum processor architectures ultimately rely on a high-fidelity method to read out the spin of each qubit. For spin readout via spin-to-charge conversion, either the energy dependent tunnelling rate to a reservoir \cite{elzerman2004single}, or between two dots for Pauli spin blockade \cite{nakajima2017robust}, should significantly exceed the spin relaxation rates. Preliminary measurements indicate that the tunnel rates in this device are too low for spin readout \cite{petta2005pulsed}. In the simplest configuration, the transition rate between (1,0) and (0,1) is measured to be $2.0\pm0.2$~Hz. Increasing the electron population so that the electronic wavefunctions for each dot overlap more strongly, shows an increase in the transition rate to $40\pm4$~Hz for the (2,0)-(1,1) charge transition. We note that transition rates of this magnitude are strongly impacted by stochastic charge movement caused by phonons or high-amplitude low-frequency electric noise.

An additional device that incorporates a metal layer (M1) approximately $270$~nm above the active silicon nanowire region was also tested (data not shown). This ‘global’ top gate can be used to tune coupling between dots by applying a large bias voltage V$_{\mathrm{M1}}$ and a compensating bias on V$_{\mathrm{RL}}$ and V$_{\mathrm{L1}}$, V$_{\mathrm{L2}}$. Initial measurements testing this mode of operation were undertaken by sweeping the L1/L2 lock-in frequency to observe the frequency at which the SET signal falls. This drop in SET signal occurs when the lock-in frequency increases beyond the inter-dot charge transition rate. For the (1,0)-(0,1) transition with V$_{\mathrm{M1}} = 9$~V, the frequency was measured to be $8.5$~kHz, which is considered still insufficient for spin readout. We also underscore that due to the floating couplers, which are responsible for forming the C1 and C2 quantum dots, changes in the V$_{\mathrm{M1}}$ bias have a strong impact on the quantum dot electrostatics and require a recalibration of the charge stability diagram. 

\begin{figure*} [ht]
	\centering
	\includegraphics[width=\linewidth]{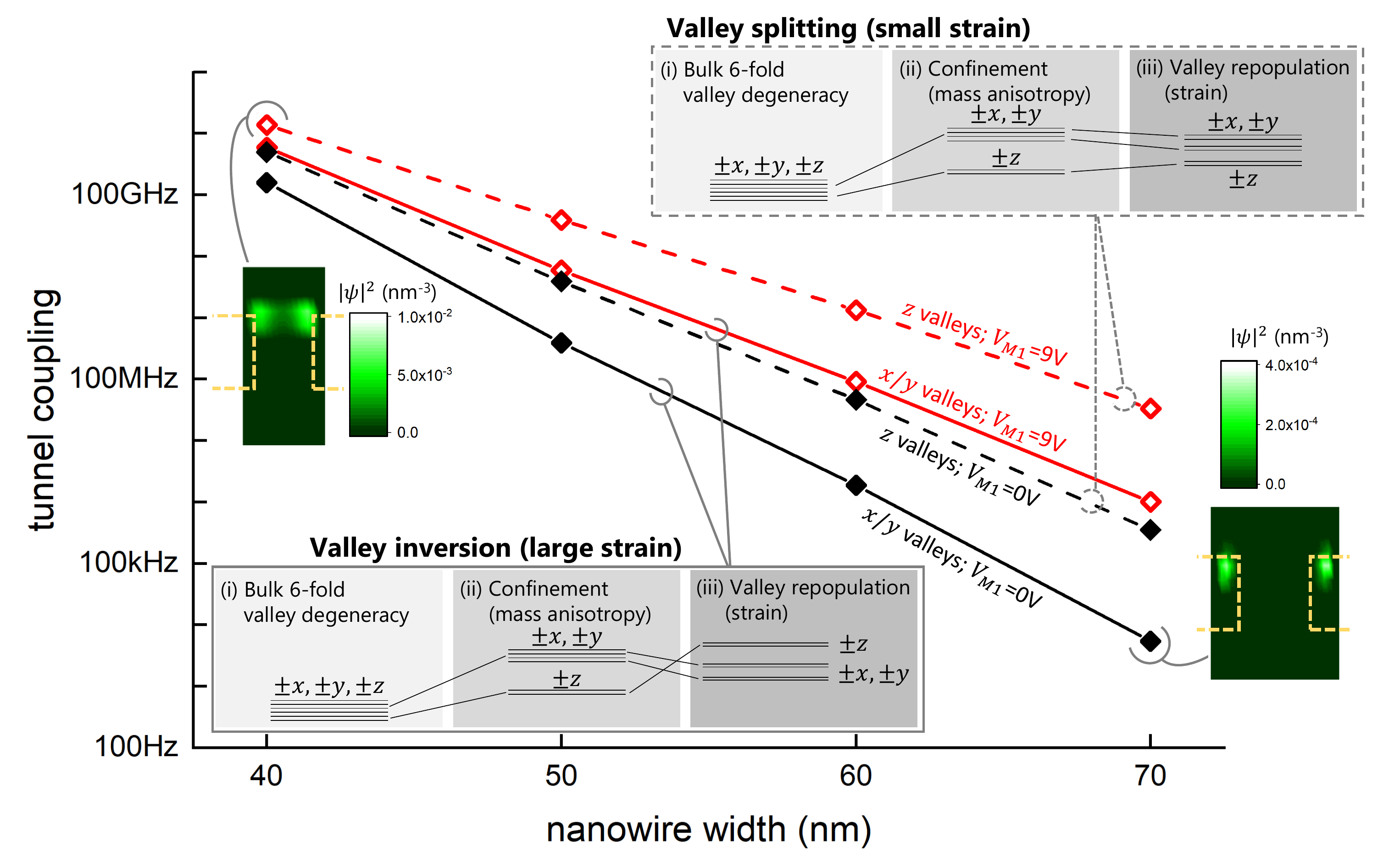}
	\caption{\textbf{Valley structure of the silicon nanowire and calculated tunnel rates.} 
	The solid lines show the tunnel rates in a nanowire in which the valley state is either $x$ or $y$, which could be caused by large strain or because $z$ valley states are occupied by a closed shell of electrons.  The corresponding inset (bottom) shows schematically (i) the six-fold degeneracy in bulk, cubically symmetric silicon, (ii) the effect of electrostatic confinement at the corner dots, which isolates the $z$ valleys as the ground state (the electric fields are slightly more vertical than a perfectly symmetric corner), and (iii) the valley repopulation into $x$ or $y$ due to the strain anisotropy. In this case, the wavefunctions extend less towards the centre of the nanowire, the electronic densities are not symmetrically distributed around the gates, and resulting tunnel rates are smaller than in the case of a $z$ valley state, shown as dashed lines, which is the ground state in the case where strain is not strongly anisotropic (see also top inset). The black lines with solid symbols show the case where the M1 metallic top gate is grounded, while the red lines with open symbols indicate the case of a biased top gate with V$_{\mathrm{M1}} = 9$~V. The electronic density distributions can be seen in the insets, comparing the limiting cases of the largest and smallest tunnel rates, which correspond to a $z$ and a $x$ ground state, respectively. The density distributions are offset with respect to the gates due to the effect of the relative negative bias on the adjacent gates.
	}\label{simulation}
\end{figure*}

\section*{ELECTRONIC STRUCTURE AND TUNNEL RATE CALCULATIONS} 

In order to design devices with tunnel rates that can enable the spin readout fidelities required for quantum computation, we model the electronic structure of nanowires of different transverse dimension within the effective mass approximation (see also Ref.~\cite{defranceschi2016soi}). Electrons in silicon possess an anisotropic effective mass. In order to determine the effective mass of the electrons in the direction of the tunnelling movement between dots, it is necessary to determine which silicon conduction-band valley state is the ground state for these quantum dots. The nanowire is fabricated along the [110] direction on a (001) wafer, however the valley states are aligned along [100] ($x$), [010] ($y$) and [001] ($z$) directions. For the geometry of the device studied here, strain and electrostatic confinement have comparable effects in determining the valley ground state (see insets in Figure~\ref{simulation}). On one hand, the corner electric field confines the electron slightly more strongly against the upper (001) oxide interface, which energetically favours a ground state formed by the $z$ valley states. On the other hand, strain due to the different thermal contractions of the material stack \cite{thorbeck2015formation} may alter the energy ordering between valley states \cite{herring1956transport}. 

Simulations show that both effects can be quantitatively comparable for all nanowire sizes - see Methods for details of the electrostatic and strain simulations, and the finite differences method for the single-particle Schroedinger equation with non-diagonal effective mass tensor. The angle of the corner electric field leads to a splitting between $z$ valleys and $x$ or $y$ valleys induced by quantum confinement of  $\Delta_{\mathrm{conf}} \approx 19-22$~meV, favouring a doubly degenerate $z$ ground state (see also Ref.~\cite{voisin2014few}). Valley-orbit coupling could lift the remaining degeneracy between $\pm z$ valleys, but it is not included in the simulations presented here. The impact of strain on the valley structure is hard to predict with precision in nanometric devices, which may be affected by process-induced strains, crystal defects, variability and non-idealities in their geometries. A strain anisotropy of $0.2\%$ would be sufficient to generate a relative energy shift between valleys that is comparable to $\Delta_{\mathrm{conf}}$ \cite{herring1956transport, fischetti1996band}. For this reason, we study the tunnel rates both with and without valley inversion.

With valley inversion, the mass along the direction connecting both L1 and C1 dots [$1\bar{1}0$] is a combination of the longitudinal and transversal masses and the effective mass tensor becomes non-diagonal. As a result, the electronic wavefunctions are not symmetric and the tunnel rates are smaller than in the case of a $z$ valley state (see Figure~\ref{simulation}). The calculations shown in Figure~\ref{simulation} indicate tunnel rates in this device become more suitable for quantum computing applications for a modest decrease in the transverse nanowire dimension to below 60~nm. The range of tuneability expected from biasing the M1 global top gate is also shown (Figure~\ref{simulation} open symbols).

\section*{DISCUSSION}

In general, any form of orbital or valley degeneracy is detrimental for quantum computing, leading to fast spin relaxation and undermining exchange coupling between spins. More specifically in the context of spin qubit readout, this added degree of freedom may hinder spin blockade. In order to guarantee a priori that no undesirable degeneracy is present, it is necessary to know the electronic structure of the dot, which can be a challenging issue for many-electron qubits. In specific examples of very small, highly symmetric quantum dots \cite{leon2020coherent}, it is possible to recognise the shell structure of the dot and consider electronic interactions as a small perturbation, similarly to what is done in atomic physics. In a more general case such as the corner dots in a nanowire, the electronic structure and excitation spectrum may not bear an easily recognisable labelling in terms of quantum numbers. In these conditions, it is desirable to operate at low electron occupancies.

The charge sensor measurements described here enable the identification of quantum dot transitions down to the last electron. High sensitivity charge sensing is a key ingredient used to perform single-shot readout of single spin qubits \cite{nakajima2017robust}. Nevertheless, the device studied here does not allow for spin readout due to the limited tunnel rates between dots. In addition, the source and drain leads are not ideal reservoirs for spin readout due to their intrinsic inhomogeneity resulting from the doping profile. Our study provides a pathway for future device designs that could reach the tunnel rate regime required for spin readout based on Pauli spin blockade.

\section*{METHODS}

\subsection*{Experimental techniques} The devices measured here were fabricated using 300~mm FD-SOI technology \cite{vinet2018towards}. The 70~nm nanowire is etched to align with the [110] direction and a 6~nm thermal oxide is grown for gate insulation. The gates and floating couplers consist of 6~nm of TiN underneath 50~nm of heavily doped poly-Si, patterned with electron-beam lithography to 40~nm width and 40~nm separation. A large (30~nm) silicon nitride spacer is deposited and etched to self-align the phosphorus and arsenic source-drain implant to the gates and to minimise the probability of dopants entering the channel during the implant process. The implants are activated with a rapid thermal anneal. Standard CMOS back-end of line processing was used to connect the device through to the bond pads for assembly.
After assembly, devices were first measured at $T= 4.2$~K by dipping into liquid helium and checking for expected turn-on characteristics, including regular Coulomb blockade oscillations in the source-drain current. Pending a successful $T= 4.2$~K measurement, the devices were mounted for measurement in a closed-cycle dilution refrigerator with a base temperature of $T= 20$~mK. There was no intentional magnetic field applied during the experiment.

The general connection configuration for measurement is shown in Figure~\ref{device}b. DC voltages were supplied by Stanford Research Systems SIM928 isolated low-noise voltage sources through a 5:1 voltage divider. A Stanford Research Systems SR830 DSP-based lock-in amplifier was used to apply AC excitations to the source-drain bias and measure transport current in a single nanowire. A second SR830 applies AC excitation to dot gates L1 and L2 to measure inter-nanowire transconductance via the current measured in the SET sensor nanowire for detection of charge movements induced by the excitation on L1 and L2. Sensitivity of the SET charge-sensor is kept constant over large sweeps using the feedback mechanism as described in \cite{yang2011dynamically}. The SET current signal is amplified at room temperature with a FEMTO DLPCA-200 transimpedance amplifier then fed into a Stanford Research Systems SIM910 JFET pre-amplifier for ground isolation. Real time traces were acquired using a Siglent SDS2204X oscilloscope. 
Details on the measurement and extraction of the tunnel rates can be found in the Supplementary Material.

\subsection*{Simulation techniques} The electrostatic potential was calculated numerically using the Poisson solver within COMSOL. The geometry was input as-designed and the bias on the gates was taken from the experimental values corresponding to the middle of the L1-C1 (0,1)-(1,0) anti-crossing. The dot gate and coupler were held at the same potential for simplicity. When simulating narrower nanowire widths as in Figure~\ref{simulation}, the gate separation was reduced by 5~nm for every 10~nm reduction in nanowire width.

The potential extracted from the numerical simulation was then input into a bespoke 3D anisotropic effective mass solver. A binary search algorithm made small adjustments to the field in the [$1\bar{1}0$] direction separating the dots in order to find the minimum eigenvalue separation and the (0,1)-(1,0) tunnel coupling. We note that the measured tunnel rate is not necessarily equivalent to the tunnel coupling since inelastic charge relaxation processes might play a role at such low transition rates, for instance through phonon emission \cite{wang2013charge}. Further details on the simulation method, including information on the valley structure of the quantum dots, can be found in the Supplementary Material.

\section*{ACKNOWLEDGMENTS}
We acknowledge support from Silicon Quantum Computing P/L, the Australian Research Council (FL190100167, CE170100012 and LE160100069), the NSW Node of the Australian National Fabrication Facility and UNSW Sydney. W.G. acknowledges an Australian Government Research Training Program Scholarship.

\bibliographystyle{naturemag}
\bibliography{refs}

\clearpage
\renewcommand{\tablename}{\textbf{Supplementary Table}}
\renewcommand\thetable{\textbf{\Roman{table}}}

\beginsupplement

\section*{Supplementary Information}
\subsection*{Measurement of wrap-around-gate nanowire devices}

A double-nanowire device with wrap-around gates and similar dimensions to that presented in Figure~\ref{device}a was also measured, a schematic of which is seen in Figure~\ref{wrapgate}a. The right nanowire is used as an SET, but with more regular and controllable Coulomb oscillations as a result of having independent control of both gates R1 and R2, as shown in Figure~\ref{wrapgate}b. The SET is used to sense charge movements in quantum dots under L1 and L2, potentially down to the last electron, see Figure~\ref{wrapgate}c. There are signs of multiple dots forming under a single gate, shown by the irregularity in the charging energy of the dots, whilst still maintaining a similar lever arm. This is particularly apparent in the close parallel transitions around V$_{\mathrm{L1}} = 0.25$~V. The issue of multiple dots under a single gate may be mitigated or eliminated with a narrower nanowire. In this device, the coupling from L1 and L2 quantum dots to the SET is weak, most likely due to the wrap-around gates shielding the electric field. This results in a significantly reduced readout signal. Some of the signal is recovered in the many electron regime as in Figure~\ref{wrapgate}d, potentially due to the quantum dots becoming large enough for some of their wave functions to extend beyond the edges of the top gates. 

\begin{suppfig*} [h]
	\centering
	\includegraphics[width=\linewidth]{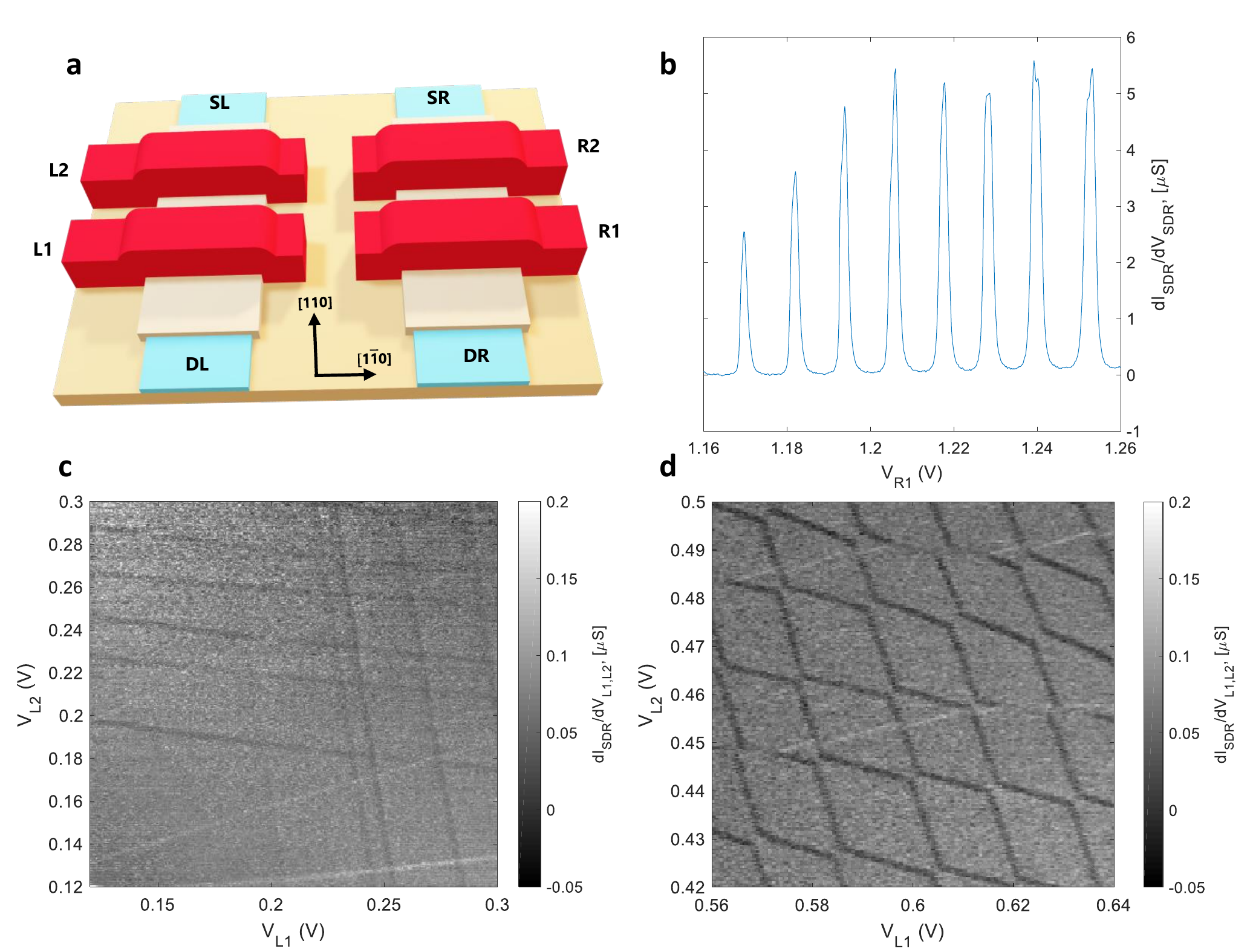}
	\caption{\textbf{Measurement of wrap-around-gate nanowire devices.} \textbf{\lowercase{a}}, Schematic of a double-nanowire device with wrap-around gates. \textbf{\lowercase{b}}, Coulomb oscillations of the sensor-nanowire. \textbf{\lowercase{c}}, Charge stability map of the dot-nanowire in the few electron regime. \textbf{\lowercase{d}}, Charge stability map of the dot-nanowire in the many electron regime.
	}\label{wrapgate}
\end{suppfig*}

\subsection*{Measurement of tunnel rates}

Interdot tunnel rates were measured in two devices; device \#1 as presented in the main text, and another of identical design, device \#2. Changes in tunnel rates due to changes in electron occupancy are observed via two measurement techniques. In device \#1, a real-time trace of sensor current is measured at the charge anti-crossing of the double dot L1-C1, as seen in Figures~\ref{tunnelratemeas}a,b. Here we measure transition rates of 2~Hz and 48~Hz for the (1,0)-(0,1) and (2,0)-(1,1) charge transitions, respectively. This measurement provides some indication of tunnel rates, but is highly dependent on the effective electron temperature, and noise profile of the dots. 

Further to this, in device \#2 tunnel rates are measured by driving across the charge anti-crossing at a frequency f$_{\mathrm{probe}}$, then sweeping f$_{\mathrm{probe}}$ to observe the decay in read-out signal as f$_{\mathrm{probe}}$ surpasses the inter-dot tunnel rate, seen in Figures~\ref{tunnelratemeas}c-e. This technique reveals a significantly higher tunnel rate than that measured via real-time spectra, but indicates a similar trend of increased tunnel rates for higher dot occupancies, with $8.5$~kHz for 1 electron, increasing to $>100$~kHz for 4 electrons. A slight decrease to $3.3$~kHz is seen for 2 electrons, possibly constrained by a spin relaxation effect and Pauli spin-blockade.

\begin{suppfig*} [h]
	\centering
	\includegraphics[width=\linewidth]{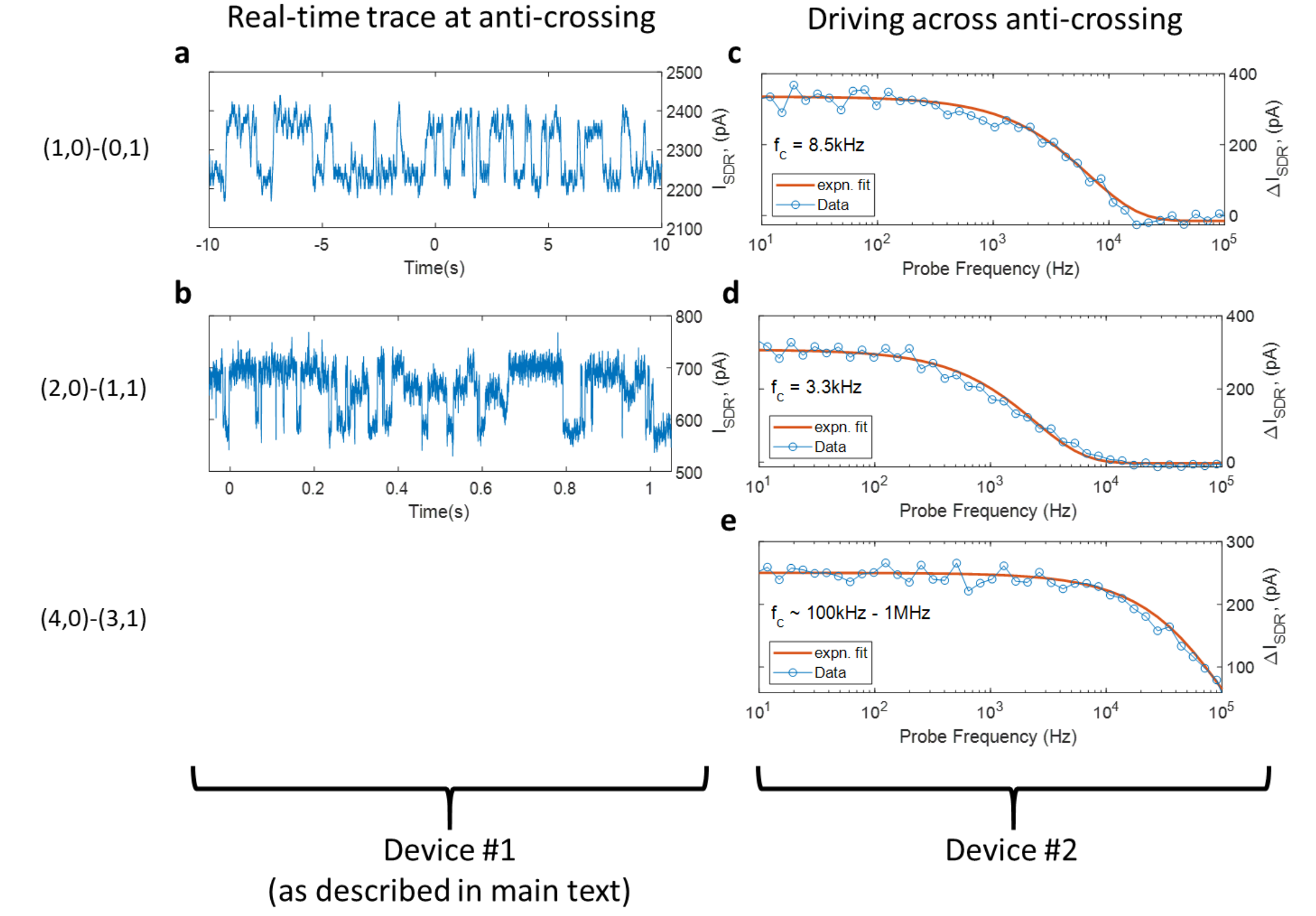}
	\caption{\textbf{Tunnel rate measurements.} \textbf{\lowercase{a,b}}, Real-time current traces of charge movements across the (1,0)-(0,1) and (2,0)-(1,1) anti-crossings. \textbf{\lowercase{c-e}}, Direct measurement of driven charge movements across the (1,0)-(0,1), (2,0)-(1,1) and (4,0)-(3,1) anti-crossings w.r.t. drive frequency, measured in a second device of the same design.
	}\label{tunnelratemeas}
\end{suppfig*}

\subsection*{Electrostatic potential and strain simulations}

Electrostatic potential numerical modelling was carried out using the Poisson solver within COMSOL. The geometry was generated using designed parameters and the gate potentials in the simulation were taken from the experimental gate biases that set the double dot to the middle of the (0,1)-(1,0) charge transition line. An example of the resulting potential distribution is shown in Figure~\ref{strainsims}a, representing a slice through the middle of the nanowire thickness, parallel to the plane of the buried oxide [(001) plane]. The potential of the couplers was kept the same as the adjacent dot gate for simplicity. Results from the electrostatic potential simulations were fed into a multivalley effective mass simulation code to calculate the tunnel coupling between adjacent dots. 

Strain was also simulated in COMSOL on the same geometry using a linear elastic material model and anisotropic elasticity tensor \cite{hopcroft2010what, thorbeck2015formation}. We adopted the room temperature values for the coefficient of thermal expansion of the constituent materials, which are expected to give an upper bound on strain from cryogenic cooling. The elastic constants for all materials were taken to be isotropic, except for the silicon nanowire, which is considered to be a single crystal. Three strain components are plotted in Figure~\ref{strainsims}b, with x, y and z being the coordinates parallel to the sides of the nanowire (which corresponds to the [110], [$1\bar{1}0$] and [001] crystallographic directions, respectively). We note that shear strain can be comparable to the strain along the principal axes, and even though it is not plotted here we also consider its influence on the electronic structure of the dots. The temperature range simulated was 423~K (150~${\degree}$C) to 1~K, designed to capture an effect from the raised process temperature. Full process-induced strains were not included here.

\begin{suppfig*} [h]
	\centering
	\includegraphics[width=\linewidth]{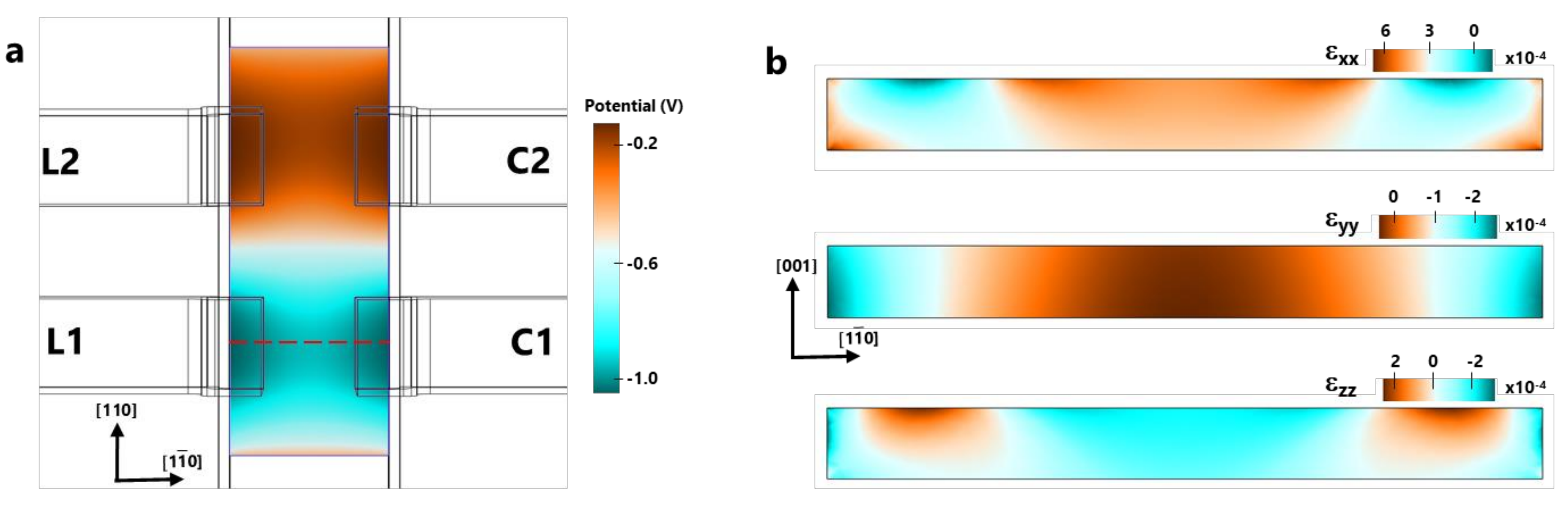}
	\caption{\textbf{Electrostatic potential and strain simulations.} \textbf{\lowercase{a}}, Electrostatic potential calculated using COMSOL showing a cut at half the thickness of the nanowire. \textbf{\lowercase{b}}, Elastic strain components plotted on the cross-section across the nanowire (as marked in \textbf{a}).
	}\label{strainsims}
\end{suppfig*}

\subsection*{Electronic Structure and Multivalley Effective Mass Simulations}

The conduction band of pristine silicon has six degenerate minima at values of the crystalline momentum vector $k=(0.84\times \frac{2π}{a_0})$\^{u}, where $a_0=0.543$~nm is the lattice parameter of silicon and \^{u} are unit vectors pointing along the $\pm x$,$\pm y$, and $\pm z$ directions as defined by the conventional cubic lattice. The dispersion relation around each of these minima is anisotropic, giving rise to an effective mass tensor described by two parameters – the longitudinal effective mass $m_l=0.98m_0$, which describes the curvature along the direction of $\pm$\^{u}, and the transversal mass $m_t=0.19m_0$, which describes the axially symmetric dispersion in any direction perpendicular to $\pm$\^{u}.

The device geometry and material stack modify this valley structure by breaking the cubic symmetry. Three main effects are discussed here: The electrostatic potential, strain and the interface induced valley-orbit coupling.

Firstly, the split gate geometry creates electric fields at an angle with regard to the corners of the nanowire. Depending on the particular angle of this field, the ground state electronic wavefunction might be bound more tightly against the upper (001) surface or against the ($1\bar{1}0$) surface at the side of the nanowire. Since the effective mass of the electron is largest along the direction of the valley, the details of this confinement angle will lead to a difference in confinement energies for each valley state. An energetic advantage for $\pm z$ valleys is expected for a potential more tightly confined against the (001) nanowire upper interface, and $\pm x$ and $\pm y$ valleys for the ($1\bar{1}0$) side interface. In all our simulations, we observe an energy difference between 19~meV and 22~meV between these valley states.

The cubic symmetry of the crystal is further broken by the anisotropic strain fields within the device. This is a result of the difference in thermal expansion coefficients of the various materials in the stack. This effect is hard to model, since some strain may be in the materials even at room temperature, as created by the fabrication process steps. Moreover, stacking faults, dislocations, voids and geometric imperfections, even if very far from the active region of the nanowire, generate a propagated strain field that is hard to predict. Our idealised model calculations indicate that the energy shifts of the different valleys amount to a few meV, which means that in our system the strain field inhomogeneity can have an impact on the details of the quantum dot formation, but that the overall anisotropy of the strain tensor is not large enough to create an inversion of the valley quantum number of the ground state. 

Both effects mentioned above isolate the $\pm z$ valleys as the ground state, which at this point is doubly degenerate. The degeneracy between $+z$ and $-z$ would be lifted by valley-orbit coupling generated by the (001) upper interface of the nanowire. For this reason, a sharp flat oxide interface is desirable. This will impact spin qubit properties, such as relaxation time, exchange coupling and singlet-triplet blockade. In our current analysis, this valley-orbit splitting is not considered. Valley interference could impact the tunnel coupling estimated here, for instance if the interface does not remain flat from one quantum dot to another. This level of detailed information is not observable from our experiments, so that our multivalley effective mass model disregards valley-orbit coupling.

\clearpage

\end{document}